\documentstyle[twocolumn,aps]{revtex}
\input{psfig.sty}

\begin{document}
\draft
\title{Topological phases and circulating states of Bose-Einstein condensates}
\author{K. G. Petrosyan and L. You}
\address{School of Physics, Georgia Institute of Technology,
Atlanta, GA 30332-0430}
\date{\today}
\maketitle

\begin{abstract}
We show that the quantum topological effect predicted by Aharonov and Casher
(AC effect) [Phys. Rev. Lett. {\bf 53}, 319 (1984)] may be used to create
circulating states of magnetically trapped atomic Bose-Einstein condensates (BEC).
A simple experimental setup is suggested based on a multiply connected
geometry such as a toroidal trap or a magnetic trap pinched by a blue-detuned
laser. We give numerical estimates of such effects within the current atomic
BEC experiments, and point out some interesting properties of the associated
quantized circulating states.
\end{abstract}

\narrowtext

Theoretical and experimental research has advanced rapidly since the
first experimental evidence of the atomic Bose-Einstein condensation
in alkali metal gases \cite{bec,mit-1,bec-2}.
Compared with strongly interacting $^4$He systems,
dilute atomic quantum gases opened an unique opportunity
to perform detailed comparative studies of the predictions of the
many body theory \cite{edwards}. In the last several years, experimentalists
have demonstrated remarkable success testing theoretical concepts and
calculations based on the theory of weakly interacting bose gases developed
many years ago \cite{edwards}. However, somewhat surprisingly,
the realization and the detection of a vortex state \cite{kac,fetter}
has not been demonstrated despite the
tremendous efforts from several leading experimental groups \cite{v1}.

Apart from potentially nontrivial physical reasons preventing the
direct observation of the vortex state, we believe the lack of
robust and effective creation schemes has been a major factor.
Several theoretical groups have recently proposed ideas
for vortex creation. Walsworth {\it et. al} \cite{ron} proposed
a  magnetic dipole Franck-Condon coupling scheme.
Marzlin {\it et. al} \cite{zhang} and Bolda {\it et. al} \cite{walls}
have recognized that the circulations of photon mode functions can be
coupled into the condensate through the electric dipole
Franck-Condon factor. Dum {\it et. al} \cite{dum} developed
a robust scheme based on adiabatic following.
Rokhsar \cite{rokhsar} and Javanainen \cite{juha} have considered
interesting properties of the formation and behavior of
vortex/circulating states. Goldstein {\it et. al} \cite{meystre}
proposed a possible detection scheme.
The interesting connection with superfluidity and persistent current states
in a toroidal shaped geometry was addressed in \cite{bloch,mueller}.

In this paper we propose a new scheme for the creation of quantized
circulating states \cite{fetter1}. Our main idea involves the use of
the Aharonov-Casher (AC) effect \cite{casher},
the magnetic ``dual'' of the Aharonov-Bohm (AB) effect \cite{bohm}.
Our discussion is based on the setup as illustrated in Fig. 1,
a 3D magnetically trapped condensate is pierced along the cylindrically
symmetric z-axis by a far blue detuned laser as used in the first MIT
BEC experiment \cite{mit-1}. We envision that there exists an imbedded
conducting wire inside the laser beam which can be charged to
produce a radially
directed electric field. Such a setup is reminiscent of the standard
discussion of the AB/AC effects \cite{bohm,casher}.
We also assume that the atoms are spin polarized by the trap
field along the axial direction $\hat z$. Such a setup may
require the development of new types of magnetic traps.
The AC effect to be utilized here has been extensively discussed in
many papers \cite{cyclic,ulf}, and will not
be the detailed subject of this paper.

We note that the non-relativistic effective single atom
Hamiltonian takes the form \cite{casher},
\begin{equation}
H=\frac{1}{2M}(\vec p-\vec E \times \vec \sigma/c)^2 -
\frac{\sigma^2 E^2}{Mc^2}+ V(\vec r),
\label{h1}
\end{equation}
where $M$ is the mass of the atom. $\vec E$ is the static
electric field strength. $\vec \sigma$ denotes the magnetic
dipole moment of the trapped atom and $c$ is the vacuum speed
of light. The trapping potential contains two parts:
the static magnetic trap $V_{t}(\vec r)$ and the
optical dipole potential from the piercing laser field $V_{L}(\vec r)$.
The second quantized form for atoms described by (\ref{h1}) is
\begin{eqnarray}
{\cal H}=\int d\vec r\left[
\hat \psi^{\dag} (\vec r)(H-\mu) \hat \psi(\vec r)
+\frac{u_0}{2} \hat \psi^{\dag 2}(\vec r)\hat \psi^2(\vec r)\right].
\label{h2}
\end{eqnarray}
We have introduced the chemical potential $\mu$ to guarantee the
conservation of total number of atoms. We have also used the
shape independent form for the atom-atom interaction with
$u_0={4\pi \hbar^2 a_{\rm sc}}/M$ where $a_{\rm sc}$ is the
s-wave scattering length.

For alkali metal atoms under these conditions the atomic spin is
$\vec \sigma = g_F\mu_B \hat z$,
(where typically the Lande $g_F$ factor is of order one).
$\mu_B={\hbar e}/{2m_ec}$ is the Bohr magneton for the electron
(with mass $m_e$). Assuming the charged wire to be long and thin
compared with all other dimensions in the system as illustrated
in Fig. 1 (a), we approximate the
electric field with the infinite wire limit
$\vec E(\vec \rho)=2n_e \hat \rho/\rho$, where $n_e$ is the linear
charge density along the wire. This implies that $\vec E \times \vec \sigma$
is in the azimuthal direction $\hat\phi$,
effectively acting as a torque term. Such a term
will induce a non-trivial ground state of condensed atoms.

At zero temperature ($T=0$) we introduce an order parameter
for the condensate $\psi(\vec r)=\langle \hat\psi(\vec r)\rangle$
(normalized to one).
Minimizing the Hamiltonian (\ref{h2}) with respect to
$\psi(\vec r)$, we obtain the standard nonlinear Schr\"odinger equation
\begin{eqnarray}
&&\left[-\frac{\hbar^2}{2M}\left(\frac{1}{\rho}\frac{\partial}{\partial
\rho}
\rho \frac{\partial}{\partial \rho}  +
\frac{\partial^2}{\partial z^2} \right) \right. \nonumber\\
&&-\frac{\hbar^2}{2M\rho^2}\left(\frac{\partial}{\partial
\phi}-i\eta\right)^2
  + Nu_0 |\psi(\rho,\phi,z)|^2 + \nonumber\\
&&\left. V(\rho,\phi,z)-\frac{\hbar^2}{M}\frac{\eta^2}{\rho^2}
\right]
 \psi(\rho,\phi,z)=\mu\,\psi(\rho,\phi,z),
\label{nlse3d}
\end{eqnarray}
where $N$ is the total number of condensed atoms.
We define the dimensionless term $\eta=2n_e g_F\mu_B /c\hbar$.
Typical numerical values are estimated as
\begin{eqnarray}
\eta= 2\times {N_e e\over {\hbar\over m_e c}}\times g_F\times {e\hbar\over 2 m_e c}
\times {1\over c\hbar}
=N_e g_F \alpha,
\end{eqnarray}
where we have expressed the linear charge density in terms of
the number of electronic charge units $N_e$ per Compton length
$^-\!\!\!\!\lambda_c={\hbar/ m_e c}\approx 3.862\times 10^{-13}{\rm m}$.
$\alpha$ is the fine structure constant: $e^2/\hbar c\approx1/137$.
For the AC effect to be observable, one would
require $\eta\sim 1$, which is equivalent to a linear
charge density of about
\begin{eqnarray}
n_e \sim 1/(g_F\alpha ^-\!\!\!\!\lambda_c)
\sim 3.55\times 10^{14}/g_F\, ({\rm m}^{-1})
\end{eqnarray}
or larger, i.e. a charge distribution of at least several
electronic charges per Compton length. Such a charge distribution
would create a large electric field in the trapping area.
At a distance of $\bar\rho\, (^-\!\!\!\!\lambda_c)$ from the wire
(where $\bar\rho$ is $\rho$ in units of $^-\!\!\!\!\lambda_c$),
one can estimate the radial electric field to be
\begin{eqnarray}
E(\vec \rho)&\sim& 2\times N_e e/(\bar\rho\, ^-\!\!\!\!\lambda_c^2)\nonumber\\
&\sim& 2\times (e/a_0^2)\,
(a_0^2/\, ^-\!\!\!\!\lambda_c^2)/(\alpha g_F\bar\rho) \nonumber\\
&\sim& 5\times 10^{6}/(g_F\bar\rho)\ ({\rm a.u.}),
\end{eqnarray}
about $(2/g_F)\times 10^{-3}\ ({\rm a.u.})$
at a distance of $1$ (mm) away from the wire. A toroidal trap
with a radius $\sim 1$ (mm) as illustrated in Fig. 1 (b) may
make these conditions accessible.

In the case of a toroidal trap we consider an alternative charge
arrangement consisting of a charged sphere instead of an infinitely
long wire [see Fig. 1 (b)]. The electric field in the plane of the torus
can then be conveniently expressed as $N_e e/\rho_0^2$ with $\rho_0$ the
radius of the torus. Assuming the width
of the torus tube to be much smaller than its radius, one can effectively
find that
\begin{eqnarray}
\eta={N_e e\over \bar\rho_0 {\hbar\over m_e c}}
\times g_F\times {e\hbar\over 2 m_e c}
\times {1\over c\hbar}
=N_e g_F \alpha/2\bar\rho_0.
\end{eqnarray}
For $\eta$ to be of order of $1$ one would require
\begin{eqnarray}
N_e\sim 2\bar\rho_0/g_F \alpha.
\end{eqnarray}
We estimate the magnitude of the electric field inside
the torus tube to be
\begin{eqnarray}
E(\vec \rho_0)&\sim& {N_e e\over \rho_0^2}
={2\over \alpha g_F}\times {1\over \bar\rho_0}
\times \left ({a_0\over ^-\!\!\!\!\lambda_c}\right)^2
\times {e\over a_0^2}\nonumber\\
&=&{2\over \alpha g_F}\times {1\over \bar\rho_0}
\times 1.885 \times 10^4\ ({\rm a.u.}),
\end{eqnarray}
$a_0=\hbar^2/m_e e^2$ is the Bohr radius.
For a torus with a radius of about 1 mm,
the electric field is about $(5/g_F) 10^{-4}$ (a.u.).
We note the atomic unit (a.u.) for the field strength
is $5.142 \times 10^{9}$ (V/cm). One may hope that such
field strength could be realized in the future \cite{note}.

Assuming the toroidal trap can confine the condensate to a width
much narrower than the radius $\rho_0$, we approximate
the three dimensional problem as describe above in Eq. (\ref{nlse3d})
by an effective one dimensional one along the azimuthal direction
$\phi$. To continue our analysis, we take the following ansatz for
the transverse structure of the toroidally confined condensate.
\begin{eqnarray}
\psi(\rho,\phi,z)&=& \Phi(\rho,z) \psi(\phi)/\sqrt{\rho_0},\nonumber\\
\Phi(\rho,z)&=&{\frac{1}{\sqrt{{\cal N}}}}\,\exp \left[
-{\frac{1}{4}}\left( \frac{(\rho-\rho_0)^2}{\sigma_{\rho}^{2}}+\frac{z^{2}}
{\sigma_{z}^{2}}\right) \right] ,
\end{eqnarray}
with the normalization
${\cal N}=2\pi\,\sigma_{\rho} \sigma_{z}$. $\sigma_{\rho}$ and
$\sigma_{z}$ are the effective ground state width in $\rho$ and $z$
respectively.

The effective 1D equation of motion for $\psi(\phi)$ is
\begin{eqnarray}
\left[ -\left(\frac{\partial}{\partial\phi}-i\eta\right)^2
+\tilde{u}_{0} \left|\psi(\phi)\right|^2\right]\psi(\phi)
=\mu_{\rm eff}\psi(\phi),
\label{nlse1d}
\end{eqnarray}
where
\begin{eqnarray}
\tilde{u}_{0} &=& Nu_0\times
{1\over 4\pi \rho_0^2\,\sigma_{\rho} \sigma_{z}}
\Big/\left(\frac{\hbar^{2}}{2M \rho_0^2}\right),\nonumber\\
\mu_{\rm eff} &=& \frac{\eta^2}{2}+\tilde{u}_{0}
+\langle V(\rho,\phi,z)\rangle
\Big/\left(\frac{\hbar^{2}}{2M \rho_0^2}\right) \nonumber\\
&&-\rho_0^2\left\langle\frac{1}{\rho}\frac{\partial}{\partial \rho}
\rho \frac{\partial}{\partial \rho}+
\frac{\partial^2}{\partial z^2} \right\rangle.
\end{eqnarray}
$\langle V(\rho,\phi,z)\rangle$ is the potential averaged over the
transverse profile $\Phi(\rho,z)/\sqrt{\rho_0}$. It is
constant since the original potential $V(\rho,\phi,z)$ is
azimuthally symmetric and independent of $\phi$.
The last term of $\mu_{\rm eff}$ is the transverse kinetic energy,
and can be approximated as
\begin{eqnarray}
\left\langle\frac{1}{\rho}\frac{\partial}{\partial \rho}
\rho \frac{\partial}{\partial \rho}+
\frac{\partial^2}{\partial z^2}\right\rangle
\approx
\left\langle\frac{\partial^2}{\partial \rho^2}+
\frac{\partial^2}{\partial z^2}\right\rangle,
\end{eqnarray}
over the state $\Phi(\rho,z)/\sqrt{\rho_0}$. This is
also constant, determined by the detailed transverse profile
of the trapping potential.

Now it is easy to appreciate the consequences of the
topological phase term $\eta$ as described by the equation
(\ref{nlse1d}). Thermodynamically the ground state should
be the state with the minimum value of $\mu_{\rm eff}$.
Because of the required
periodic boundary condition $\psi(\phi)=\psi(\phi+2\pi)$, the
uniform state in $\phi$ has to be of the form $\psi(\phi)\sim e^{im\phi}$,
where $m$ is the integer of quantized circulation, or the angular momentum
along the $z$-axis. For such a state one obtains
\begin{eqnarray}
\mu_{\rm eff}=(m-\eta)^2+{\tilde{u}_{0}\over 2\pi}.
\end{eqnarray}
Minimizing the above with respect to $m$ we obtain that $m=[\eta]$,
where $[\eta]$ denotes the nearest integer of $\eta$. Since $\eta$
can be controlled experimentally, one may expect the $m$ value
of the circulating state to jump whenever $\eta$ crosses half integer
values. This result is illustrated in Fig. (\ref{fig2}) as solid steps
for $T=0$. We note that the dot-dashed line for the case of
$0<T<T_c$ (the condensation temperature) is simply a
weighted average of the result for $T=0$ with that of the thermal atoms
(denoted by the dotted line).
This result is similar to the Figures 3 and 4 of Ref. \cite{tony}
except now there is no rotation of the trap.

We can also check the local stability of the above circulating states.
In the case of a superposition state
\begin{eqnarray}
\psi(\phi)={1\over \sqrt{2\pi}}
[\sqrt{1-x}\,e^{im\phi}+\sqrt{x}\,e^{i\theta} e^{i(m+1)\phi}],
\end{eqnarray}
where $x$ is a variational mixing parameter and $\theta$ a random
phase factor. One then finds
\begin{eqnarray}
\mu_{\rm eff}&=&(1-x)\,(m-\eta)^2+x\,(m+1-\eta)^2\nonumber\\
&&+{\tilde u_0\over 2\pi} [1+2x(1-x)].
\end{eqnarray}
Since $\mu_{\rm eff}=(m-\eta)^2+{\tilde u_0\over 2\pi}$
at $x=0$ ($m$ state) and
$\mu_{\rm eff}=(m+1-\eta)^2+{\tilde u_0\over 2\pi}$
at $x=1$ ($m+1$ state), we see that the lower
energy state is indeed determined by the value $[\eta]$.
The curves connecting x-values are inverted parabolas
(for $\tilde u_0 >0$) peaked at $x={1\over 2}+(m+{1\over 2}-\eta)/
{\tilde u_0\over \pi}$ where $\mu_{\rm eff}$ takes
the value of
\begin{eqnarray}
\mu_{\rm eff}&=& \left(1+{\pi\over \tilde u_0}\right)
(m-\eta)(m+1-\eta)\nonumber\\
&&+{1\over 2}\left(1+{\pi\over 2\tilde u_0}+{3\tilde u_0\over 2\pi}\right).
\end{eqnarray}
To reach the lower side of $[\eta]$ from $m$ to $m+1$ or vise versa,
the potential barrier in between has to be overcome. In the present
case, this is achieved first by adiabatically climbing the
potential barrier to the top when $\eta$ is experimentally adjusted,
and then sliding down to the new minimum of $\mu_{\rm eff}$.
This is illustrated in Fig. \ref{fig3}.

An important practical concern for the AC effect as applied
here is the effect of electric field to polarize atoms.
As previously shown by several authors, a polarized
atom inside a crossed electric and magnetic field experiences
additional AB/AC like topological effects \cite{ulf}. In particular,
the effective contribution is of the same form as discussed
above, but
\begin{eqnarray}
\eta_{E\times B}=\alpha(0)n_e B/(2\hbar c),
\end{eqnarray}
for a linear charge distribution. $B$ is the magnetic
trapping field in the $\hat z$ direction and
$\alpha(0)=\bar\alpha(0) (a_0^3)$ is the polarizability; with
$\bar\alpha(0)$ typically of the order of a few hundred for
alkali metal atoms. In dimensionless units one obtains
\begin{eqnarray}
\eta_{E\times B} &=&\bar\alpha(0)\times a_0^3 \times {N_e e\over a_0}
\times B\times {1\over 2\hbar c}\nonumber\\
&=&\bar\alpha(0)\times N_e \left ({\mu_B B \Big/{e^2 \over a_0}}\right),
\end{eqnarray}
where $N_e$ now is the number of charges per Bohr radius.
The last term in parenthesis is the
Bohr magneton interaction energy with the external field,
divided by the atomic unit for energy. It
is $\sim 2 \times 10^{-10}$/(gauss).
With typical values for $N_e$, $\bar\alpha(0)$, and $B$,
we conclude that the same topological effect due to the applied
electric field and the magnetic trapping field is much smaller,
$\eta_{E\times B}\ll 1$.

In conclusion, we have estimated the Aharonov-Casher effects on
the quantum degenerate bose gas. We have shown that due to
the presence of a topological phase term, the ground state of
a degenerate quantum gas will display quantized flux states, i.e.
circulating states of definite angular momentum. Our numerical
estimates have shown that they would be difficult to
realize within the current BEC experiments, but are potentially
observable in the future. We want to emphasize
that although our model is based on an effectively toroidal
shaped trap, the topological effects discussed require only
a multiply connected geometry, i.e. independent of the details of
the transverse shape of the condensate profile.
Therefore the results as illustrated in Figs. 2-3 should be
qualitatively valid even when a more rigorous approach is taken.
Our proposal makes the crucial first link between the topological
phases and atomic Bose-Einstein condensates, and can also be
further explored along the direction of using Bose Gases as a
system in which the AC effect can be realized.

We would like to thank M. R. Andrews for helpful discussions.
This work is supported by the NSF grant No. PHY-9722410.
We thank A. Idrizbegovic and G. Stark for help in making the
figures, and Mr. Stark for proofing.
L. Y. wants to thank all BEC workshop participants
for several organized discussion sessions on vortex
states at the Institute for Theoretical Physics (ITP), Santa Barbara.
He would also like to thank the ITP for its hospitality and the support
of the NSF grant No. PHY94-07194.

\begin{figure}
\caption{Two possible setups for vortex/circulating state
creation with AC effects. (a) magnetic trap pinched by a
blue detuned laser with an imbedded charged line;
(b) a toroidal trap surrounding a charged sphere distribution.
}
\label{fig1}
\end{figure}

\begin{figure}
\caption{The ground state index $m$, i.e. the angular momentum per atom
at zero temperature ($T=0$). The dotted line is the result for a
classical (noncondensed) ensemble of atoms whose individual atomic
angular momentum is not quantized. The tilted dot-dashed line is
the result for $T\ne 0$ but $<T_c$. }
\label{fig2}
\end{figure}

\begin{figure}
\caption{A plot of $\mu_{\rm eff}$ showing the local stability
of the $m=[\eta]$ states in the free energy functional space.
We have taken the parameter $\tilde u_0/2\pi=2$.}
\label{fig3}
\end{figure}


\begin{references}
\bibitem{fetter1}A certain amount of confusion exists in the current
literature on the meaning of a vortex state. We will use the
term circulating state for an angular momentum eigenstate
in a multiply connected geometry such as a torus discussed here, and
vortex state to denote those eigenstates in a simple connected
geometry, i.e. supported by only the nonlinear interaction term.
L.Y. thanks A. Fetter for clarifying this confusion to him.
\bibitem{note}For a spherical conductor with a radius
of $\sim$ 500 ($\mu$m), one has to charge to $\sim$ 30 (kV)
in order to realize the required electric field at a distance of
1 (mm). Use of Dielectric materials will lower the charging voltage.
\end{references}
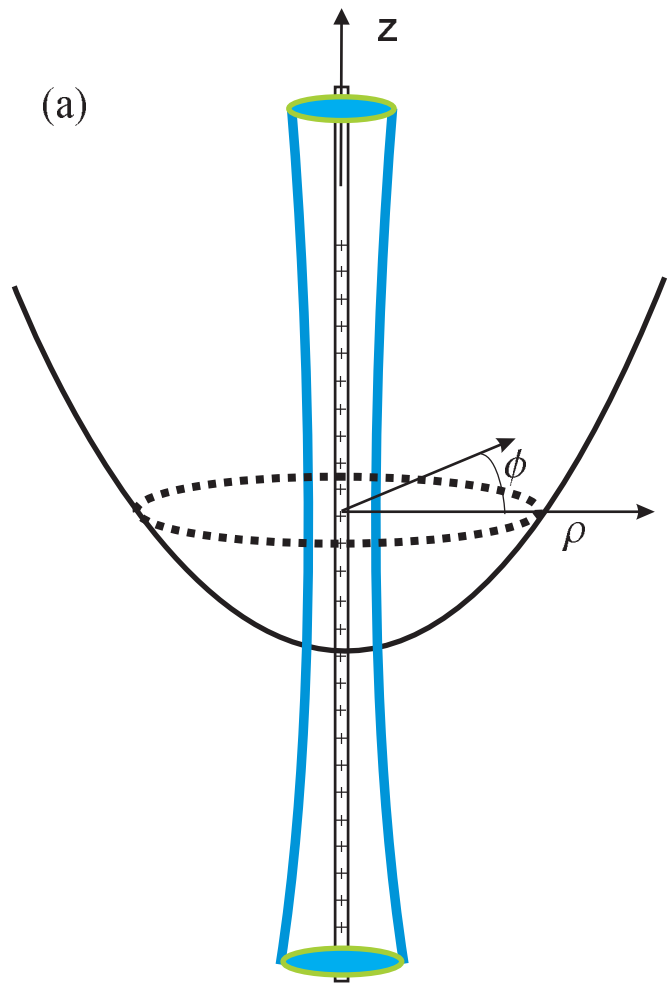
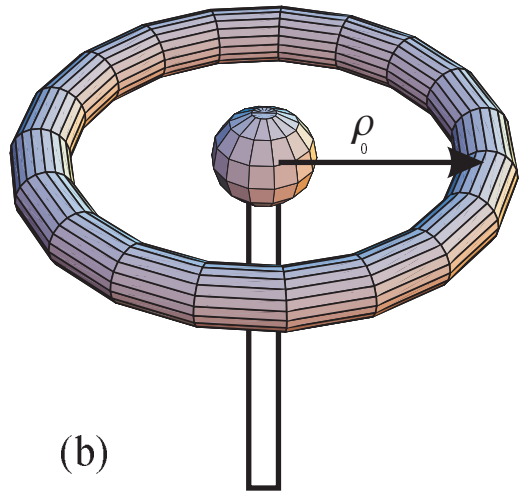
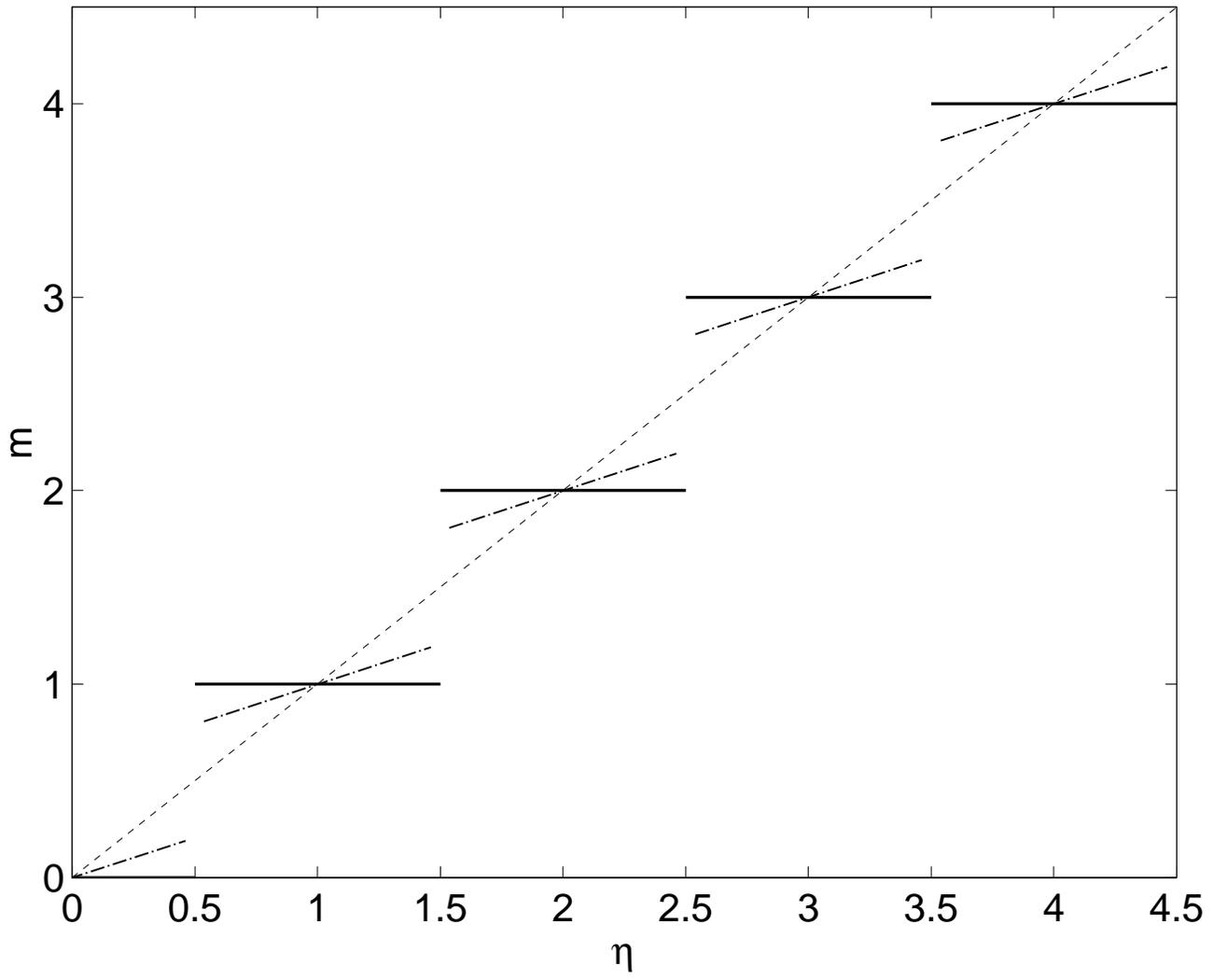
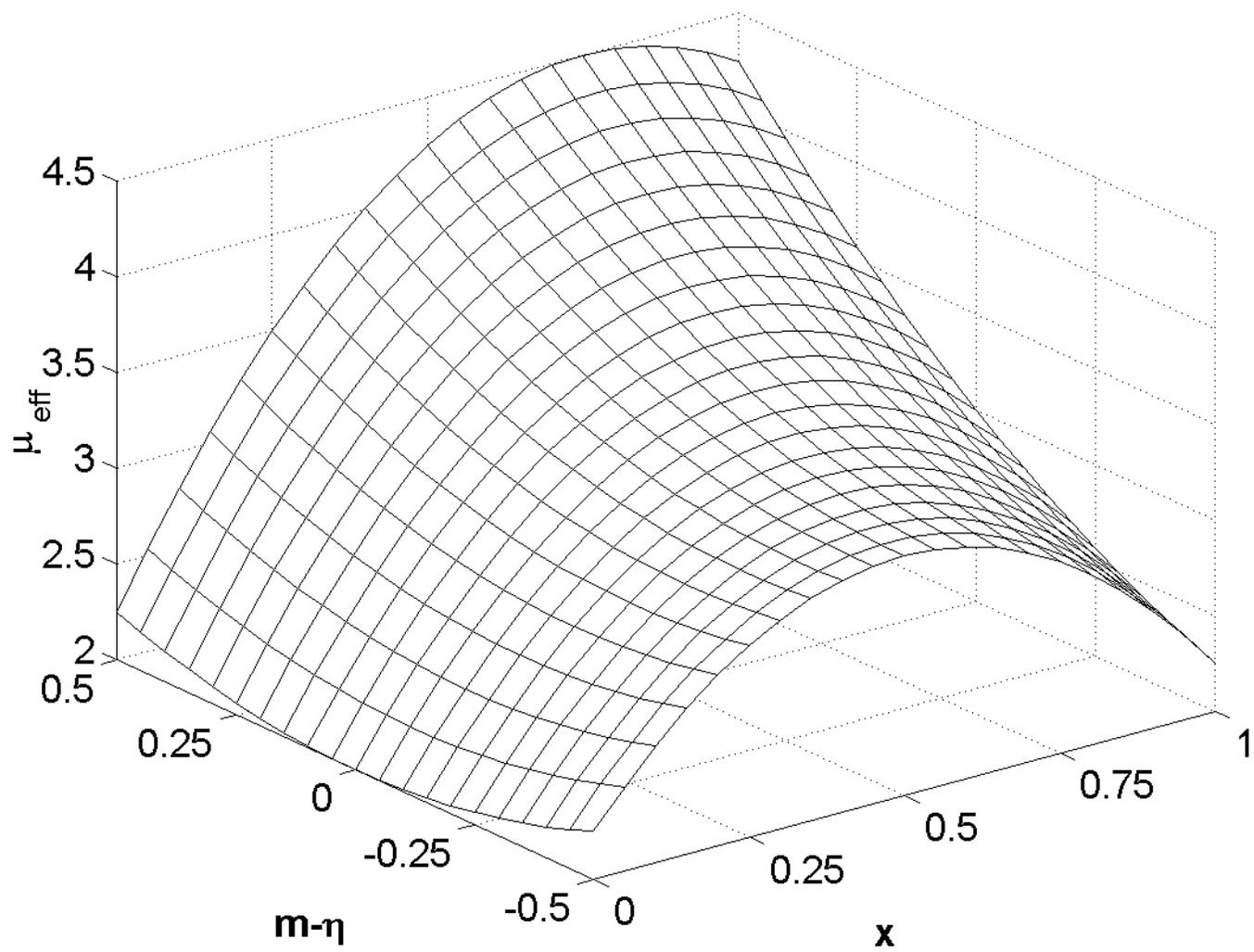
\end{document}